\documentclass[a4paper,11pt]{article}
\usepackage{jheppub} % for details on the use of the package, please see the JINST-author-manual
\usepackage{amsmath,amssymb,mathtools,bm}
\usepackage{physics}
\usepackage{braket}
\usepackage{xcolor}
\preprint{YITP-26-57}

\arxivnumber{2606.03671} 

\makeatletter
\renewcommand{\@fpheader}{\phantom{Prepared for submission to JHEP}}
\makeatother

\title{\boldmath 
Quantum Matter Makes Lightcones Quantum
}

\author[a,b]{Tomohiro Fujita}
\affiliation[a]{Department of Physics, Ochanomizu University,
2-1-1 Otsuka, Bunkyo-ku, Tokyo 112-8610, Japan}
\affiliation[b]{Kavli Institute for the Physics and Mathematics of the Universe (WPI), The University of
Tokyo Institutes for Advanced Study, The University of Tokyo, Chiba 277-8583, Japan}

\author[b,c,d,e,f]{and Misao Sasaki}
\affiliation[c]{Asia Pacific Center for Theoretical Physics, Pohang 37673, Korea}
\affiliation[d]{Center for Gravitational Physics, Yukawa Institute for Theoretical Physics, Kyoto University, Kyoto 606-8502, Japan}
\affiliation[e]{Leung Center for Cosmology and Particle Astrophysics, National Taiwan University, Taipei 10617, Taiwan}
\affiliation[f]{Department of Physics, Pohang University of Science and Technology (POSTECH)
Pohang 37673, Korea}
\emailAdd{fujta.tomohiro@ocha.ac.jp}

\abstract{
In gravitational physics, matter does not merely move within spacetime; it also determines the light cones that define causal relations.  What happens when the matter that determines these light cones is itself in a quantum state?  We address this question in a controlled low-energy setting: a massless scalar field propagating in the spacetime with the Newtonian gravitational potential sourced by a non-relativistic quantum particle.
We show that the light cones are affected by an operator-valued Shapiro delay, with the three consequences: 
(i) causal-boundary shifts are promoted to noncommuting observables, giving the causal structure an irreducible quantum uncertainty;
(ii) the causal relation between two fixed spacetime points can become a superposition of timelike and spacelike configurations;  
and (iii) tracing out the source smears the Wightman light-cone singularity, producing an effective UV cutoff.
Thus, quantum matter does not merely fluctuate within spacetime; it makes the causal structure itself quantum, even without including propagating graviton modes.
}
\begin{document}
\maketitle
\flushbottom

%%%%%%%%%%%%%%%%%%%%%%%%%%%%%%%%%%%%%%%%%%%%%%%%%%%
\section{Introduction}
\label{sec:intro}
%%%%%%%%%%%%%%%%%%%%%%%%%%%%%%%%%%%%%%%%%%%%%%%%%%%

A complete quantum theory of gravity remains unknown. 
At the same time, the quantum nature of matter is beyond doubt. 
This simple fact already raises a conservative question: how does the gravitational field inherit quantum properties from quantum matter?
This question can already be asked in the low-energy, nonrelativistic regime, where propagating gravitational degrees of freedom are negligible, while the matter source itself must still be treated quantum mechanically.

Various approaches have addressed related questions. 
Quantum field theory (QFT) in curved spacetime has been remarkably successful as a framework for quantum fields on a fixed classical geometry \cite{Birrell:1982ix,Wald:1995yp,Parker:2009uva}. 
Primordial cosmology provides one of its most prominent applications, where quantum fluctuations of fields on an expanding background are stretched to cosmic scales and become the seeds of structures observed today~\cite{Sasaki:1986hm,Mukhanov:1988jd,Planck:2018jri}. 
Gravitational waves are also often quantized around a fixed background~\cite{Starobinsky:1979ty}. 
Moreover, semiclassical gravity and stochastic gravity have sought approximate ways of incorporating the influence of quantum matter on gravity~\cite{Page:1981aj, Diosi:1984wuz,Penrose:1996cv,Hu:2008rga,Bahrami:2014gwa}. 
However, these frameworks either assume a classical background geometry or describe its quantum effects only approximately; they are still different from treating the geometry as a quantum object. 
%In contrast, we address the problem more directly: we keep the matter source itself quantum, and ask how the causal structure probed by a relativistic quantum field is affected.

Here we study this question in a minimal weak-field setting. 
The source is a single nonrelativistic quantum particle, and the probe is a real massless scalar field propagating in the spacetime with a weak Newtonian gravitational field generated by the source. 
This setup is simple enough to allow explicit field-theoretic calculations, while still retaining the essential feature that the gravitational source is quantum. 
The Newtonian regime has also recently attracted renewed attention through proposals for experimental tests of gravity-induced entanglement~\cite{Bose:2017nin,Marletto:2017kzi,Fujita:2023pia}. 
Here we instead ask how such a quantum source affects the light cones probed by a relativistic field. 
The key quantity is the Shapiro time delay~\cite{Shapiro:1964uw,Will:2014kxa}: once the source position is an operator, the Newtonian potential and hence the Shapiro delay become operator-valued. 
Thus the light cone itself acquires an operator-valued shift. 
We show that this leads to three consequences: (i) noncommuting causal-boundary shifts, (ii) branch-dependent timelike/spacelike causal relations, and (iii) smearing of the Wightman light-cone singularity after tracing out the source.

The topics discussed in this paper have important precedents, although in rather different contexts. 
From the perspective of Planck-scale quantum gravity, spacetime uncertainty relations and noncommutative spacetime have long suggested that spacetime localization may be subject to irreducible quantum limitations~\cite{Doplicher:1994tu,Doplicher:1994zv}. 
In a different, more operational direction, indefinite causal order has been studied in the quantum information context, where even the order of events may fail to be fixed~\cite{Oreshkov:2011er,Zych:2017tau}.
Light-cone fluctuations and the smearing of Wightman singularities were studied in the context of quantized gravitons~\cite{Ford:1994cr,Ford:1996qc,Matsui:2026ppg}. 
These works provide important conceptual precedents. 
The assumptions and frameworks, however, are distinct from ours. 
In much of the indefinite-causal-order literature, for example, the emphasis is operational and quantum-information-theoretic.
By contrast, our approach derives the quantum causal structure through an explicit calculation within the standard framework of relativistic QFT.
What is striking is that these ingredients, which have appeared in different lines of thought, arise together as natural consequences of one simple quantity: the quantum Shapiro delay.

The main purpose of this paper is to give a minimal and explicit field-theoretic realization of quantum causal structure induced by quantum matter. 
Our result does not amount to a complete theory of quantum gravity, nor does it assume the quantization of propagating gravitational degrees of freedom. 
Rather, it shows that the classical-background picture already becomes insufficient, in a precise sense, once the matter sourcing the geometry is treated quantum mechanically. 

This paper is organized as follows. 
In Sec.~\ref{sec:setup} we introduce our setup. 
In Sec.~\ref{sec:operator_valued_causal_boundary} we derive the quantum Shapiro delay. 
In Sec.~\ref{sec:noncommuting_causal_boundary_shifts} we show that causal-boundary shifts are promoted to noncommuting observables. 
In Sec.~\ref{sec:causal_relation_superposition} we consider a source particle in a spatial superposition and show that the causal relation can become branch-dependent. 
In Sec.~\ref{sec:reduced_wightman_smearing} we discuss the smearing of the Wightman light-cone singularity. 
We summarize and discuss implications in Sec.~\ref{sec:sum_dic}.

%%%%%%%%%%%%%%%%%%%%%%%%%%%%%%%%%%%%%%%%%%%%%%%%%%%
\section{Setup: the source particle and the massless scalar field}
\label{sec:setup}
%%%%%%%%%%%%%%%%%%%%%%%%%%%%%%%%%%%%%%%%%%%%%%%%%%%

We consider a real massless scalar field propagating in a weak gravitational field in $3+1$ dimensions.
The gravitational field is sourced by a nonrelativistic quantum particle, and our goal is to understand how the quantum-ness of the source particle is imprinted on the causal structure probed by the scalar field.
Throughout this paper we use natural units, $c=\hbar=1$.

We take the source to be a nonrelativistic particle of mass $M$.
Its position and momentum operators are denoted by $\hat{\bm q}$ and $\hat{\bm p}$, and its Hamiltonian is
\begin{equation}
    \hat H_S
    =
    \frac{\hat{\bm p}^{\,2}}{2M}
    +
    V(\hat{\bm q}) \,,
    \label{eq:source_Hamiltonian}
\end{equation}
where $V(\hat{\bm q})$ is the potential.
We assume that the source is sufficiently heavy and that its motion remains nonrelativistic.
We next specify the weak gravitational field generated by this quantum source.
We restrict attention to the quasistatic Newtonian regime, neglecting propagating gravitational degrees of freedom and retardation effects.
In Newtonian gauge, the corresponding weak-field metric is written as
\begin{equation}
    ds^2
    =
    -\left(1+2\hat\Phi_g\right)dt^2
    +
    \left(1-2\hat\Phi_g\right)
    \delta_{ij}dx^i dx^j .
    \label{eq:metric_weak_field}
\end{equation}
Here $\hat\Phi_g$ is not an independent  degree of freedom, but the constrained gravitational potential determined by the source density.
In the Newtonian regime, the Hamiltonian constraint takes the form of the Poisson equation,
\begin{equation}
    \nabla^2 \hat\Phi_g(\bm x)
    =
    4\pi G M\delta^{(3)}\!\left(\bm x-\hat{\bm q}\right),
    \label{eq:operator_Poisson}
\end{equation}
where $G$ denotes Newton's gravitational constant.
Solving this equation with asymptotically flat boundary conditions gives
\begin{equation}
    \hat\Phi_g(\bm x)
    =
    -\frac{GM}{|\bm x-\hat{\bm q}|}\,.
    \label{eq:operator_potential_schrodinger}
\end{equation}
Thus, the gravitational potential becomes an operator on the source Hilbert space because the position of the source is quantum mechanical.
\footnote{
Our treatment should not be confused with semiclassical gravity, in which the metric is sourced by the expectation value of the matter stress tensor,
$G_{\mu\nu}=8\pi G\langle\hat T_{\mu\nu}\rangle$.
In the Newtonian regime, the corresponding equation would be
$\nabla^2\Phi_g=4\pi G\langle\hat\rho_S\rangle$,
and $\Phi_g$ would remain a classical $c$-number field.
}
In the following, we assume
\begin{equation}
    |\hat\Phi_g|\ll 1
\end{equation}
and keep only the leading contribution in perturbation theory in $\hat\Phi_g$

We now introduce a real massless scalar field $\phi$ propagating in this weak gravitational field.
We assume that the scalar field is in the free-field vacuum $\ket{0}_\phi$ in the asymptotic past.
The gravitational field sourced by the scalar field itself is higher order in the perturbative expansion and will be neglected.
In this sense, the scalar field is used as a probe of the weak gravitational field generated by the quantum source.
The scalar action is
$S_\phi
    =
    -\frac12
    \int \dd t\,\dd^3x\,
    \sqrt{-g}\,
    g^{\mu\nu}
    \partial_\mu\phi\,\partial_\nu\phi$ as usual.
Substituting the metric \eqref{eq:metric_weak_field} and keeping terms up to first order in $\hat\Phi_g$, we obtain
\begin{equation}
    S_\phi
    =
    \frac12
    \int \dd t\,\dd^3x\,
    \left[
        \left(1-4\hat \Phi_g\right)\dot\phi^2
        -
        \left(\bm\nabla\phi\right)^2
    \right]
    +
    \mathcal{O}(\Phi_g^2) .
    \label{eq:scalar_action_weak_field}
\end{equation}
Here the canonical momentum is $\Pi=(1-4\hat\Phi_g)\dot{\phi}$.
Canonical quantization is imposed in the usual way,
\begin{equation}
    \left[
        \hat\phi(\bm x),
        \hat\Pi(\bm y)
    \right]
    =
    i\delta^{(3)}(\bm x-\bm y),
    \qquad
    \left[
        \hat\phi(\bm x),
        \hat\phi(\bm y)
    \right]
    =
    \left[
        \hat\Pi(\bm x),
        \hat\Pi(\bm y)
    \right]
    =
    0 ,
    \label{eq:canonical_commutation_relations}
\end{equation}
where, for convenience, we take the Schr\"odinger picture with $\hat\phi(\bm x)$ and $\hat\Pi(\bm x)$ being the Schr\"odinger-picture operators at a reference time.
We note that the canonical variables $\hat\phi$ and $\hat\Pi$ act only on the scalar-field Hilbert space and commute with the source operators, while $\dot{\hat\phi}=(1+4\hat \Phi_g)\hat \Pi$ involves both the source and scalar-field Hilbert spaces. 

The Hamiltonian of the scalar field is
$\hat H_\phi=\hat H_0+\hat H_{\rm int}+\mathcal{O}(\Phi_g^2)$, where
\begin{equation}
    \hat H_0
    =
    \frac12
    \int \dd^3x\,
    \left[
        \hat\Pi^2(\bm x)
        +
        \left(\bm\nabla\hat\phi(\bm x)\right)^2
    \right]
    \label{eq:free_scalar_Hamiltonian}
\end{equation}
is the free Hamiltonian, and
\begin{equation}
    \hat H_{\rm int}
    =
    2
    \int \dd^3x\,
    \hat\Phi_g(\bm x)\,
    \hat\Pi^2(\bm x)
    \label{eq:interaction_Hamiltonian}
\end{equation}
is the leading interaction Hamiltonian.
The total Hamiltonian is therefore
$\hat H=\hat H_S+\hat H_0+\hat H_{\rm int}$.
We employ the interaction picture with respect to $\hat H_S+\hat H_0$, with the interaction Hamiltonian given by
\begin{equation}
    \hat H_{{\rm int},I}(t)
    =
    2
    \int \dd^3x\,
    \Phi_g\!\left(\bm x-\hat{\bm q}_I(t)\right)
    \hat\Pi_I^2(t,\bm x),
    \label{eq:interaction_picture_Hamiltonian}
\end{equation}
where
$\hat{\bm q}_I(t)
=
e^{i\hat H_S t}
\hat{\bm q}
e^{-i\hat H_S t}$
and
$\hat{\Pi}_I(t,\bm x)
=
e^{i\hat H_0 t}
\hat{\Pi}(\bm x)
e^{-i\hat H_0 t}$.
This interaction Hamiltonian is the starting point of the perturbative calculations below.

%%%%%%%%%%%%%%%%%%%%%%%%%%%%%%%%%%%%%%%%%%%%%%%%%%%
\section{Quantum Shapiro Delay}
\label{sec:operator_valued_causal_boundary}
%%%%%%%%%%%%%%%%%%%%%%%%%%%%%%%%%%%%%%%%%%%%%%%%%%%

%--------------------------------------------------
\begin{figure}[!t]
\centering
\includegraphics[width=.5\textwidth, trim=72mm 45mm 72mm 40mm,
  clip]{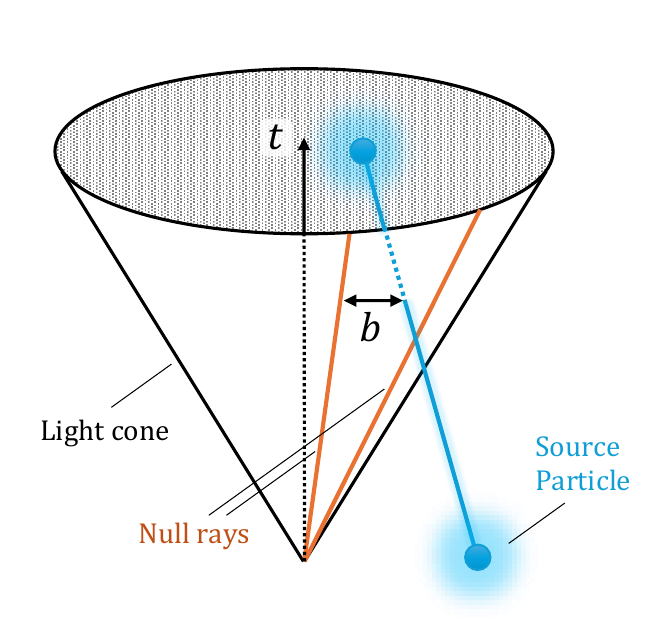}
\caption{
Schematic configuration of the unperturbed light cone (black cone) and the source-particle trajectory (blue line).  
We study how null rays (orange lines) passing near the quantum-mechanical source (blue dot) are affected by its weak Newtonian gravitational potential. 
The impact parameter $b$ is the closest spatial separation between an unperturbed ray and the source trajectory. 
The Shapiro time delay is not yet reflected in this picture.
}
\label{fig:Config}
\end{figure}
%--------------------------------------------------
In this section, we present the main result of the paper.
In flat spacetime, the direct wavefront of a massless scalar field in $3+1$ dimensions propagates on the fixed classical light cone $t=r$, as illustrated in Fig.~\ref{fig:Config}.
When the gravitational field is sourced by a quantum particle, however, the position of the light cone is shifted by an operator-valued quantity, denoted by $\hat\tau_{\bm n}$.
We first motivate this shift from the classical propagation speed of a scalar wavefront.
We then verify it by a perturbative quantum calculation, showing that the light-cone singularities of both the commutator and the Wightman function are shifted by the same operator $\hat\tau_{\bm n}$.
We start with the free propagation of the scalar field from the spacetime origin $(0,\bm 0)$ to the observation point $(t,\bm x)$, and we define
\begin{equation}
    r=|\bm x|,
    \qquad
    \bm n=\frac{\bm x}{r},
    \qquad
    u=t-r .
    \label{eq:r_n_u_def}
\end{equation}
%Here, $u$ measures the displacement from the unperturbed Minkowski light cone, $t=r$.

\begin{figure}[!t]
\centering
\includegraphics[width=.5\textwidth, trim=72mm 40mm 72mm 40mm,
  clip]{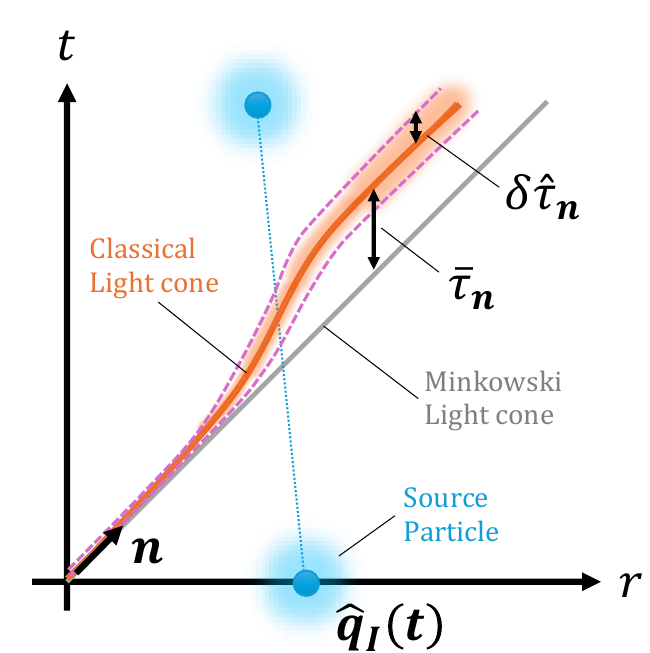}
\caption{Schematic picture of the quantum Shapiro delay $\hat{\tau}_{\bm n}$ defined in Eq.~\eqref{eq:quantum_Xi_def}. 
The blue dot represents the quantum source particle at $\bm x =\hat{\bm q}_I(t)$. 
The weak Newtonian potential sourced by this particle shifts the original Minkowski light cone (gray diagonal line). 
The orange solid line shows the mean delayed light cone, shifted by 
$\bar{\tau}_{\bm n}=\langle \hat{\tau}_{\bm n}\rangle$, while the fluctuation 
$\delta\hat{\tau}_{\bm n}=\hat{\tau}_{\bm n}-\bar{\tau}_{\bm n}$ makes its position quantum-mechanically uncertain, as indicated by the purple dashed lines. 
This is a 2D projection of the 3+1-dimensional geometry; the ray passes near the source without intersecting it (see Fig.~\ref{fig:Config}).
}
\label{fig:Lightcones}
\end{figure}

In the weak-field metric, the position of the wavefront is determined by the principal part of the scalar wave equation.
From the action \eqref{eq:scalar_action_weak_field}, for a classical potential $\Phi_g$, the principal dispersion relation is
$(1-4\Phi_g)\omega^2\simeq k^2$.
Thus the classical front velocity is
$v_{\rm front}=\omega/k\simeq 1+2\Phi_g$.
At leading order in the weak-field expansion, the bending of the ray affects the arrival time only at higher order.
Thus we evaluate the potential along the unperturbed null ray, $\bm x(t)=t \bm n$, emitted from the origin at $t=0$.
The classical arrival time is
\begin{align}
    t_{\rm arrival}
    =
    \int_0^r
    \frac{\dd t}{v_{\rm front}}
    =
    r
    -
    2\int_0^r \dd t\,
    \Phi_g\!\left( t \bm n-\bm q(t)\right)
    +
    \mathcal{O}(\Phi_g^2).
    \label{eq:classical_arrival_time}
\end{align}
The corresponding classical delay is therefore given by the integral of the gravitational potential along the unperturbed null ray,
\begin{equation}
    \tau_{\bm n}^{\rm cl}
    =
    -2\int_0^r \dd t \,
    \Phi_g\!\left( t \bm n-\bm q(t)\right) .
    \label{eq:classical_Xi}
\end{equation}
This is the leading weak-field form of the Shapiro time delay in general relativity. For a quantum source, the natural expectation is that $\bm q( t )$ in the classical expression should be replaced by the interaction-picture operator $\hat{\bm q}_I( t )$.
This gives the quantum Shapiro delay
\begin{equation}
    \hat\tau_{\bm n}
    =
    -2\int_0^r \dd t \,
    \Phi_g\!\left(
          t \bm n-\hat{\bm q}_I( t )
    \right) .
    \label{eq:quantum_Xi_def}
\end{equation}
For an attractive Newtonian potential, $\Phi_g<0$, this quantity is positive and represents a delay of the propagation.
Fig.~\ref{fig:Lightcones} illustrates a schematic picture of the quantum Shapiro delay.
%--------------------------------------------------
%--------------------------------------------------

We now confirm this expectation by perturbation theory.
To first order in the weak gravitational potential, the Heisenberg-picture scalar field is
\begin{equation}
    \hat\phi_H(t,\bm x)
    =
    \hat\phi_I(t,\bm x)
    +
    i
    \int_{-\infty}^{t}\dd t'\,
    \left[
        \hat H_{{\rm int},I}(t'),
        \hat\phi_I(t,\bm x)
    \right]
    +
    \mathcal O(\Phi_g^2),
    \label{eq:micro_3d_phiH_first_order}
\end{equation}
where
$\hat{\phi}_I(t,\bm x)=e^{i\hat H_0 t}\hat{\phi}(\bm x)e^{-i\hat H_0 t}$.
We use this expression to compute both the commutator and the Wightman function.

We first consider the unequal-time commutator,
\begin{equation}
    \Delta(t,\bm x;0,\bm 0)
    =
    \left[
        \hat\phi_H(t,\bm x),
        \hat\phi_H(0,\bm 0)
    \right].
    \label{eq:commutator_def_sec3}
\end{equation}
In flat spacetime, for $t>0$, the massless scalar commutator is
$\Delta_0(t,\bm x;0,\bm 0)=-i\delta(u)/(4\pi r)$.
The light cone is identified with the position of the delta-function singularity.
A first-order perturbative calculation gives the singular part of the commutator as (see Appendix.~\ref{app:commutator_direct_singularity} for derivation)
\begin{equation}
    \Delta_{\rm sing}(t,\bm x;0,\bm 0)
    \simeq
    -\frac{i}{4\pi r}
    \left[
        \delta(u)
        -
        \hat\tau_{\bm n}\delta'(u)
    \right]
    \simeq
    -\frac{i}{4\pi r}
    \delta\!\left(
        u-\hat\tau_{\bm n}
    \right).
    \label{eq:commutator_shift_expanded}
\end{equation}
Thus, the causal boundary is shifted from $u=0$ to $u=\hat\tau_{\bm n}$ as expected.
As discussed in Appendix~\ref{app:commutator_direct_singularity}, the terms that do not contribute to the shift of the direct wavefront are omitted here.
%
%\footnote{
%A small subtlety in the perturbative calculation is that, in addition to the $\delta'(u)$ term shown in \eqref{eq:commutator_shift_expanded}, one also finds an amplitude correction of the form
%$\Delta_{\rm sing}\supset \mathcal A_{\bm n}\delta(u)$, with $\mathcal A_{\bm n}=\mathcal O(\Phi_g)$ coming from processes similar to gravitational lensing and the integrated Sachs-Wolfe.
%This term may be interpreted as the first-order expansion of an amplitude correction carried by the delayed wavefront, $\mathcal A_{\bm n}\delta(u-\hat\tau_{\bm n})$.
%Otherwise, the classical theory would also have undelayed contributions.
%}

We next turn to the Wightman function.
Here we take the vacuum expectation value only over the scalar-field Hilbert space, leaving the source operators unevaluated:
\begin{equation}
    \hat G^+(t,\bm x;0,\bm 0)
    =
    \bra{0_\phi}
    \hat\phi_H(t,\bm x)
    \hat\phi_H(0,\bm 0)
    \ket{0_\phi}.
    \label{eq:operator_wightman_def}
\end{equation}
Near the light cone in flat spacetime, $u=t-r\simeq0$, the singular part of the free Wightman function is
$G_{0,{\rm sing}}^+(u)=-[(8\pi^2 r)(u-i\epsilon)]^{-1}$.
Including the first-order correction, the light-cone singularity becomes
(see Appendix.~\ref{app:wightman_direct_singularity} for derivation)
\begin{align}
    \hat G_{\rm sing}^+(t,r\bm n;0,\bm 0)
    \simeq
    -\frac{1}{8\pi^2 r}
    \left[
        \frac{1}{u-i\epsilon}
        +
        \frac{\hat\tau_{\bm n}}{(u-i\epsilon)^2}
    \right]
    \simeq
    -\frac{1}{8\pi^2 r}
    \frac{1}{
        u-\hat\tau_{\bm n}-i\epsilon
    } .
    \label{eq:wightman_shifted_sing_sec3}
\end{align}
Thus the light-cone singularity of the Wightman function is shifted by the same quantum Shapiro delay as the commutator.
We conclude that, in the presence of a quantum gravitational source, the location of the light cone itself does not just shift but becomes operator-valued.
In the following sections, we explore the physical consequences of this result.

%%%%%%%%%%%%%%%%%%%%%%%%%%%%%%%%%%%%%%%%%%%%%%%%%%%
\section{Noncommutativity of Causal-boundary Shifts}
\label{sec:noncommuting_causal_boundary_shifts}
%%%%%%%%%%%%%%%%%%%%%%%%%%%%%%%%%%%%%%%%%%%%%%%%%%%

In this section, we show that the operators describing the causal-boundary shifts associated with different null rays are noncommuting.
This means that the causal structure generated by a quantum source cannot be determined, as a whole, as a single sharp classical geometry.

\begin{figure}[!t]
\centering
\includegraphics[width=.5\textwidth, trim=72mm 30mm 65mm 40mm,
  clip]{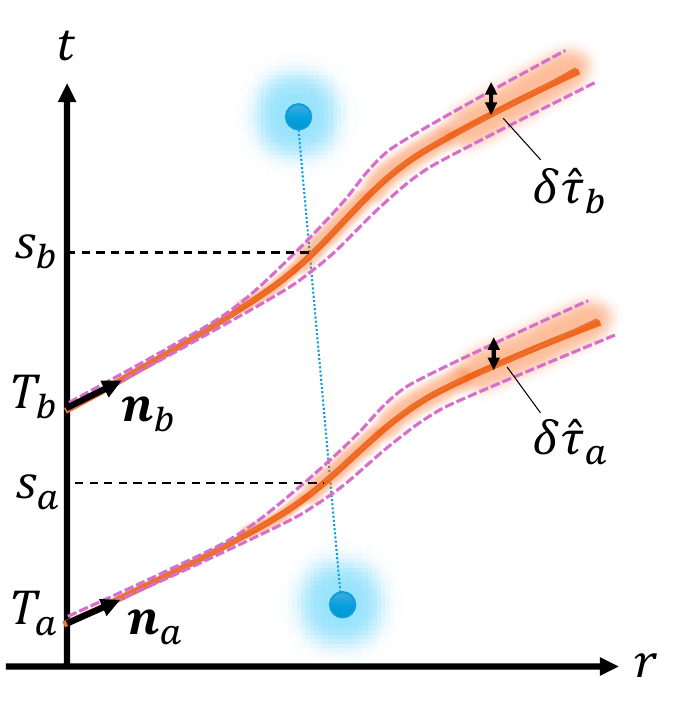}
\caption{Two null rays emitted from the spatial origin $\bm x=0$ at times $T_a$ and $T_b$ in the directions $\bm n_a$ and $\bm n_b$, respectively. 
Their quantum Shapiro-delay fluctuations, $\delta\hat{\tau}_a$ and $\delta\hat{\tau}_b$, depend on the position fluctuations of the same source particle evaluated at different times, $\delta\hat{\bm q}_I(s_a)$ and $\delta\hat{\bm q}_I(s_b)$, where $s_a$ and $s_b$ denote the respective closest-approach times. 
Since source-position operators at different times do not commute in general, the two causal-boundary shifts also do not commute with each other in general, $\left[\delta\hat\tau_a, \delta\hat\tau_b\right]\neq 0$.}
\label{fig:double_rays}
\end{figure}

So far we have used the quantum Shapiro delay $\hat\tau_{\bm n}$ for a null ray emitted from the origin at $t=0$ defined in Eq.~\eqref{eq:quantum_Xi_def}.
We now generalize this definition to an arbitrary emission time $T$.
Consider a null ray that leaves the origin $\bm x=0$ at time $T$ and reaches the observation point $\bm x=r\bm n$ at the zeroth-order arrival time $T+r$.
The corresponding quantum Shapiro delay is
\begin{equation}
    \hat\tau(T;r,\bm n)
    =
    -2
    \int_T^{T+r}
    \dd t \,
    \Phi_g\!\left(
         (t-T)\bm n-\hat{\bm q}_I(t)
    \right).
    \label{eq:quantum_Xi_with_departure_time}
\end{equation}
The associated causal boundary is then located at
$t=T+r+\hat\tau(T;r,\bm n)$.
Let two null rays be labeled by
$(T_a,r_a,\bm n_a)$ and $(T_b,r_b,\bm n_b)$, and define
$\hat\tau_a\equiv\hat\tau(T_a;r_a,\bm n_a)$ and
$\hat\tau_b\equiv\hat\tau(T_b;r_b,\bm n_b)$.
Figure~\ref{fig:double_rays} illustrates the two rays and their causal-boundary shifts.
From \eqref{eq:quantum_Xi_with_departure_time}, their commutator is
\begin{align}
    \left[
        \hat\tau_a,\hat\tau_b
    \right]
    =
    4
    \int_{T_a}^{T_a+r_a}\dd t 
    \int_{T_b}^{T_b+r_b}\dd t'\,
    \left[
        \Phi_g\!\left(
             (t-T_a)\bm n_a-\hat{\bm q}_I(t)
        \right),
        \Phi_g\!\left(
             (t'-T_b)\bm n_b-\hat{\bm q}_I(t')
        \right)
    \right].
    \label{eq:noncomm_general_commutator}
\end{align}
Notice that the above expression includes not only different spatial directions, but also the case in which the same spatial ray is emitted at different times.
It does not vanish in general, because the Heisenberg position operators of the source particle $\hat q_I$ at different times are not commuting.
Thus $\hat\tau_a$ and $\hat\tau_b$ cannot be determined simultaneously, or equivalently cannot be diagonalized simultaneously.

To see the structure more explicitly, let us consider a narrow wave packet for the source.
We denote the mean trajectory by $\bar{\bm q}(t)$ and write
\begin{equation}
    \delta\hat{\bm q}_I(t)
    =
    \hat{\bm q}_I(t)-\bar{\bm q}(t).
\end{equation}
Expanding the causal-boundary shift as
$\hat\tau_i=\bar\tau_i+\delta\hat\tau_i+\cdots$ $(i=a,b)$, the leading order fluctuation of the quantum Shapiro delay is
\begin{equation}
    \delta\hat\tau_i
    =
    2
    \int_{T_i}^{T_i+r_i}\dd t \,
    \bm\nabla\Phi_g\!\left(
         (t-T_i)\bm n_i-\bar{\bm q}(t)
    \right)
    \cdot
    \delta\hat{\bm q}_I(t).
    \label{eq:linearized_delta_Xi}
\end{equation}
%--------------------------------------------------
%--------------------------------------------------
%
%
Let us further simplify this expression in order to make the physical meaning transparent.
Assume that the source moves sufficiently slowly, and that the fluctuation of the Shapiro delay for each null ray is dominated by the neighborhood of the time at which the ray passes closest to the source.
Let $s_i$ be the time at which the ray $i\,(=a,b)$ is closest to the source particle. 
Let $b_i$ be the impact parameter at this moment, namely,
the spatial distance between the ray $i$ and the mean source position at time $s_i$ (see Fig.~\ref{fig:Config}), and let $\bm e_i$ be the unit vector pointing from the mean source position to the closest point on the ray.
Near this time, the unperturbed ray position may be written as
\begin{equation}
    (t-T_i)\bm n_i
    \simeq
    \bar{\bm q}(s_i)
    +
    b_i\bm e_i
    +
    \left(
        t-s_i
    \right)\bm n_i \quad(i=a,b).
    \label{eq:closest_approach_geometry_a}
\end{equation}
Here we have neglected the time evolution of $\bar{\bm q}$ near the closest approach, assuming that the source motion is slow compared with the propagation of light.

Using the Newtonian potential \eqref{eq:operator_potential_schrodinger}, and evaluating $\delta\hat{\bm q}_I$ at the closest-approach time $s_i$, we obtain
\begin{equation}
    \delta\hat\tau_i
    \simeq
    \frac{4GM}{b_i}\,
    \bm e_i\cdot
    \delta\hat{\bm q}_I(s_i)\quad(i=a,b).
    \label{eq:deltaXi_closest_a}
\end{equation}
Here the integration range has been sufficiently extended around the closest approach, \eqref{eq:deltaXi_closest_a} represents the asymptotic net shift after the ray has completely passed by the source.
Thus the quantum fluctuation of the Shapiro delay is dominated by the fluctuation of the source position in the direction that changes the impact parameter.

Finally, the commutator of the two Shapiro-delay fluctuations can be estimated as
\begin{equation}
    \left[
        \delta\hat\tau_a,
        \delta\hat\tau_b
    \right]
    \simeq
    \frac{16G^2M^2}{b_a b_b}\,
    e_a^i e_b^j
    \left[
        \delta\hat q_I^i(s_a),
        \delta\hat q_I^j(s_b)
    \right]
    \overset{\rm free}{=}
    16iG^2M\,\Delta T\,
    \frac{
        \bm e_a\cdot\bm e_b
    }{
        b_a b_b
    } \,,
    \label{eq:closest_Xi_commutator_final}
\end{equation}
where $\Delta T\equiv s_b-s_a$.
In the last estimate, we have approximated the source fluctuation by that of a free particle,
$\delta\hat q_I^i(t)\overset{\rm free}{=}\delta\hat q^i+\delta\hat p^i t/M$, so that
$
    \left[
        \delta\hat q_I^i(s_a),
        \delta\hat q_I^j(s_b)
    \right]
    \overset{\rm free}{=}
    i\Delta T
    \delta^{ij}/M.
$
The linear dependence on $\Delta T$ is therefore a consequence of the free-particle approximation, and should not be extrapolated to time intervals over which the potential $V(\hat q)$ appreciably modifies the source motion or the wave packet ceases to be narrow compared with $b_a$ and $b_b$.

The result \eqref{eq:closest_Xi_commutator_final} can be intuitively understood as follows.
To determine the causal-boundary shift $\hat\tau_a$ sharply is to determine sharply the transverse position
$\bm e_a\cdot\delta\hat{\bm q}_I(s_a)$
of the source at the closest-approach time of ray $a$.
By the uncertainty principle, however, such a measurement produces backreaction on the conjugate momentum component
$\bm e_a\cdot\delta\hat{\bm p}$.
After a time interval $\Delta T$, the free evolution of the particle converts this momentum uncertainty into a position uncertainty.
Therefore, if ray $b$ is later influenced the same transverse position component, its causal-boundary shift cannot be sharp at the same time.
On the other hand, if the two causal-boundary shifts depend on commuting transverse components, then $\bm e_a\cdot\bm e_b=0$, and this particular backreaction does not occur.
In this sense, the noncommutativity originates from probing the position of the same quantum source at different times.

The same noncommutativity also implies an intrinsic limitation on how sharply different causal-boundary shifts can be specified.
For a given source state, we define
\begin{equation}
    \left(
        \Delta\tau_i
    \right)^2
    \equiv
    \left\langle
        \left(
            \hat\tau_i
            -
            \left\langle
                \hat\tau_i
            \right\rangle
        \right)^2
    \right\rangle_S
    \quad
    (i=a,b) .
    \label{eq:Delta_tau_definition}
\end{equation}
The Robertson uncertainty relation then gives
\begin{equation}
    \Delta\tau_a\,\Delta\tau_b
    \ge
    \frac12
    \left|
        \left\langle
        \left[
            \hat\tau_a,
            \hat\tau_b
        \right]
        \right\rangle_S
    \right| .
    \label{eq:Xi_Robertson}
\end{equation}
Using the closest-approach estimate \eqref{eq:closest_Xi_commutator_final}, we then obtain
\begin{equation}
    \Delta\tau_b
    \ge
    \frac{
        8G^2M
    }{
        \Delta\tau_a
    }
    \left|
        \frac{
            \bm e_a\cdot\bm e_b
        }{
            b_a b_b
        }
        \Delta T
    \right| .
    \label{eq:Xi_b_uncertainty_after_Xia_measurement}
\end{equation}
Suppose that, in this state, the variance of the causal-boundary shift associated with ray $a$ is sharp,
\begin{equation}
    \Delta\tau_a\ll\Delta T\,.
    \label{eq:sigma_Xia}
\end{equation}
Then, (\refeq{eq:Xi_b_uncertainty_after_Xia_measurement}) implies that a state in which the causal boundary of ray $a$ is sharply localized cannot simultaneously make the causal boundary of ray $b$ arbitrarily sharp. Qualitatively, the dispersion $\Delta\tau_b$ is estimated as
\begin{equation}
    \Delta\tau_b\sim\frac{\Delta T}{\Delta\tau_a} \left(\frac{\ell_{\rm Pl}\,R_G}{b^2}\right)t_{\rm Pl}\,,
\end{equation}
where $\ell_{\rm Pl}=\sqrt{\hbar G/c^3}\approx1.6\times 10^{-33}$cm is the Planck length, $t_{\rm Pl}=\sqrt{\hbar G/c^5}\approx 5.4\times10^{-44}$ sec is the Planck time, 
$b$ is the mean impact parameter, and $R_G=2GM/c^2$ is the gravitational radius of the particle.
Although extremely small in the practical sense, this uncertainty is not due to a technical limitation of the measuring apparatus, but due to the quantum uncertainty principle.

If the source position were merely a classical random variable, a measurement would simply reveal, with increasing precision, an underlying classical causal structure. For a quantum source, by contrast, different causal-boundary shifts are noncommuting observables on the source Hilbert space.
Interpreting $\Delta\tau_i$ ($i=a,b$) as the precision of measurements, a measurement that sharpens one causal boundary necessarily blurs another.
Thus the causal structure generated by a quantum source is not merely a probabilistically fluctuating classical spacetime.
The quantum-mechanical uncertainty relation is inherited by the causal structure through the source matter.
Consequently, the full causal structure cannot be specified as a single sharp geometry in the same sense as in a classical background spacetime.

%%%%%%%%%%%%%%%%%%%%%%%%%%%%%%%%%%%%%%%%%%%%%%%%%%%
\section{Superposition of Causal Relations}
\label{sec:causal_relation_superposition}
%%%%%%%%%%%%%%%%%%%%%%%%%%%%%%%%%%%%%%%%%%%%%%%%%%%

We now move from the two-ray setup of the previous section to a complementary one-ray setup, in which the source particle itself is prepared in a spatial superposition.
In this setting, the different source branches produce different Shapiro delays.
We present a simple example in which a spatially delocalized source makes the causal relation between a fixed pair of spacetime points branch-dependent, placing it in a quantum superposition of timelike and spacelike configurations.

Consider a source particle in a superposition of two localized branches,
\begin{equation}
    \ket{\psi_S}
    =
    \frac{1}{\sqrt2}
    \left(
        \ket{N}
        +
        \ket{F}
    \right).
    \label{eq:source_near_far_superposition}
\end{equation}
Here $\ket{N}$ denotes the branch in which the source is close to both the null ray and the spatial origin, while $\ket{F}$ denotes the branch in which it is farther from both.
In contrast to Sec.~\ref{sec:noncommuting_causal_boundary_shifts}, where the wave-packet fluctuations of the particle played the central role, we neglect such residual position uncertainty within each branch and treat $\ket{N}$ and $\ket{F}$ as approximately localized configurations.
We also assume that the source is sufficiently heavy and approximately static on the time scale of interest.
Let $\bm e_\perp$ be a unit vector perpendicular to the propagation direction $\bm n$.
We take the source positions in the near and far branches to be
\begin{equation}
    \bm q_N
    =
    r_N \bm n
    +
    b_N\,\bm e_\perp,
    \qquad
    \bm q_F
    =
    r_F\bm n
    +
    b_F\,\bm e_\perp,
    \label{eq:qN_qF_def}
\end{equation}
and assume
\begin{equation}
    0<b_N<b_F\ll r,
    \qquad
    r_N=\frac12\left(1-\delta\right)r,
    \quad
    r_F=\frac12\left(1+\delta\right)r,
    \label{eq:bB_hierarchy}
\end{equation}
where $0<\delta<1$ controls the longitudinal separation of the two branches along the ray.
Thus, in the near branch the ray passes the source earlier and with a smaller transverse separation, whereas in the far branch it passes the source later and at a larger transverse separation.

To describe where the delay is accumulated along the ray, let $r'$ denote an intermediate distance from the origin, $0\le r'\le r$.
Using the Newtonian potential \eqref{eq:operator_potential_schrodinger}, the Shapiro delay accumulated up to $r'\bm n$ in branch $X=N,F$ is
\begin{align}
    \tau_X(r')
    =
    2GM
    \int_0^{r'}
    \frac{\dd t }{
        \sqrt{
            \left( t-r_X\right)^2
            +
            b_X^2
        }
    }
    =
    2GM
    \left[
        \operatorname{arcsinh}
        \left(
            \frac{r'-r_X}{b_X}
        \right)
        +
        \operatorname{arcsinh}
        \left(
            \frac{r_X}{b_X}
        \right)
    \right].
    \label{eq:Xi_accumulated_rprime}
\end{align}
At the final observation point $r'=r$, this becomes, in the small-impact-parameter limit,
\begin{align}
    \tau_X
    \equiv
    \tau_X(r)
    \simeq
    2GM
    \log
    \left[
        (1-\delta^2)
        \frac{r^2}{b_X^2}
    \right].
    \label{eq:Xi_impact_parameter_exact}
\end{align}
For the qualitative argument below, we again work in the small-impact-parameter limit and replace the gradual accumulation of the Shapiro delay in \eqref{eq:Xi_accumulated_rprime} by  a stepwise jump at the time when the ray is closest to the source,
\begin{equation}
    \tau_X(r')
    \simeq
    \tau_X\,\Theta(r'-r_X);
    \qquad
    X=N,F .
    \label{eq:Xi_step_approximation}
\end{equation}
In the localized-branch approximation, the quantum Shapiro-delay operator acts on the two branches as
\begin{equation}
    \hat\tau_{\bm n}(r')\ket{N}
    \simeq
    \tau_N(r')\ket{N},
    \qquad
    \hat\tau_{\bm n}(r')\ket{F}
    \simeq
    \tau_F(r')\ket{F}\,,
    \label{eq:Xi_action_NF}
\end{equation}
where $\hat\tau_{\bm n}(r')\equiv \hat \tau(0;r',\bm n)$ is defined from Eq.~\eqref{eq:quantum_Xi_with_departure_time}.

The singular part of the commutator in the presence of the quantum source is given by \eqref{eq:commutator_shift_expanded}.
We now evaluate it at an intermediate observation point $\bm x=r'\bm n$, and define $u'=t-r'$.
Acting on the superposition state \eqref{eq:source_near_far_superposition}, it gives
\begin{align}
    \Delta_{\rm sing}(t,r'\bm n;0,\bm 0)\ket{\psi_S}
    &\simeq
    -\frac{i}{4\pi r'}
    \frac{1}{\sqrt2}
    \left[
        \delta\!\left(
            u'-\tau_N(r')
        \right)
        \ket{N}
        +
        \delta\!\left(
            u'-\tau_F(r')
        \right)
        \ket{F}
    \right].
    \label{eq:comm_on_superposition}
\end{align}
In branch $X=N,F$, the causal boundary is located at $u'=\tau_X(r')$.
In the stepwise approximation \eqref{eq:Xi_step_approximation}, the near-branch delay turns on first at $r'=r_N$, while the far-branch delay turns on later at $r'=r_F$.
%Hence, the distinction is most transparent in the interval $r_N<r'<r_F$: the near branch has already acquired the shift $\tau_N(r')\simeq\tau_N$, whereas the far branch still has $\tau_F(r')\simeq0$.
In addition, since $b_N<b_F$, the final accumulated shifts satisfy $\tau_N>\tau_F$.

\begin{figure}[!t]
\centering
\includegraphics[width=.5\textwidth, trim=80mm 40mm 70mm 40mm,
  clip]{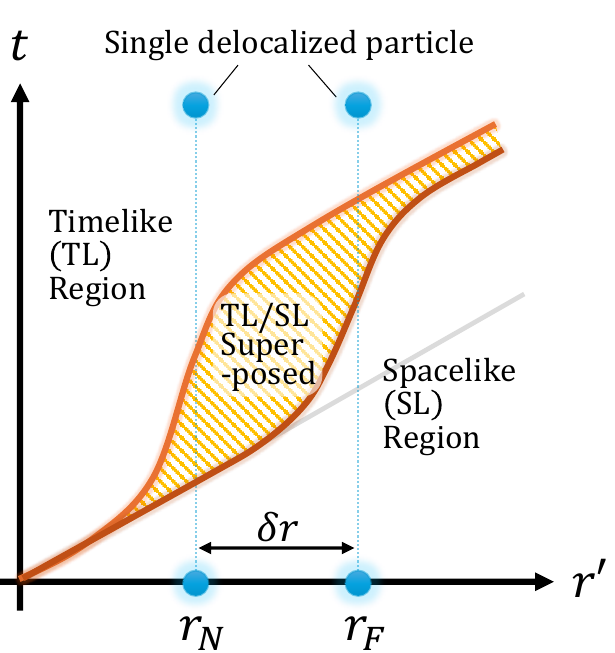}
\caption{Schematic picture of the superposition of causal relations induced 
by a single delocalized particle, namely, a single source particle in a spatial superposition.
For simplicity, the source wave packets are assumed to be sharply localized in the two branches, and their intrinsic widths are neglected. 
In the $|N\rangle$ branch (left blue dot and upper orange line) the source is closer to the ray and produces a larger Shapiro delay than in the $|F\rangle$ branch (right blue dot and lower dark orange line). 
The resulting branch-dependent causal boundaries (two orange lines) divide spacetime into three regions. 
In the intermediate region, the causal relation with the origin is timelike in one branch and spacelike in the other.}
\label{fig:superposition}
\end{figure}

Fig.~\ref{fig:superposition} illustrates how the causal structure becomes branch-dependent in this setup.
For a fixed observation distance $r'$ satisfying $\tau_N(r')>\tau_F(r')$, the two branch-dependent causal boundaries are located at $u'=\tau_N(r')$ and $u'=\tau_F(r')$, thereby dividing the spacetime points at that $r'$ into three regimes.
The intermediate region,
$\tau_F(r')<u'<\tau_N(r')$,
is the timelike/spacelike-superposed region.
In this region, the pair of coordinate points, $(0,\bm 0)$ and $(t,r'\bm n)$, is timelike in the far branch but spacelike in the near branch.%
\footnote{
Here we implicitly identify coordinate points between the branch-dependent
geometries associated with the source superposition.
In general, such an identification can be a serious issue, particularly in the strong-gravity regime. 
In the present setup, however, all branches are described as weak Newtonian
perturbations of a common asymptotically Minkowski background, written in
Newtonian gauge.
The emission and observation events are assumed to lie sufficiently far from
the source, so that its gravitational influence on the emitter and detector is negligible.
These events can therefore be identified across the branches relative to the
common asymptotic Minkowski frame.
}
Outside this interval, the causal relation is the same in both branches: spacelike for $u'<\tau_F(r')$ and timelike for $u'>\tau_N(r')$.
%
%--------------------------------------------------
%--------------------------------------------------
Note that this branch dependence does not indicate a violation of microcausality. In each branch, the lightcone singularity of the commutator lies on the corresponding causal boundary, and causality is respected within that branch.

This is in sharp contrast with a fixed classical background spacetime, where the causal relation between two points is uniquely determined by the background metric. 
Our example demonstrates that, when the gravitational source is quantum, the causal structure of spacetime cannot be defined independently of the state of that source. 
This feature is specific to gravity. 
If, for example, a charged particle were placed in a spatial superposition, the electromagnetic field sourced by it would not by itself determine the light cones, and hence would not directly change the causal relation between two spacetime points. 
Gravity is different: through the metric, it determines the light cones and therefore the causal structure. 
Thus, when the gravitational source is in a quantum superposition, the causal structure itself acquires quantum nature. 
This is a quantum consequence of the fact that gravity is the geometry of spacetime.

Before closing this section, let us estimate the size of the timelike/spacelike-superposed region. 
From Eq.~\eqref{eq:Xi_impact_parameter_exact}, its asymptotic width is
\begin{equation}
    \Delta\tau
    \equiv
    \tau_N-\tau_F
    \simeq
    4t_{\rm Pl}
    \frac{M}{m_{\rm Pl}}
    \log\!\left(
        \frac{b_F}{b_N}
    \right),
    \label{eq:Delta_tau_estimate}
\end{equation}
where $m_{\rm Pl}=\sqrt{\hbar c/G}\approx 22\,{\rm \mu g}$
is the Planck mass.
%and the Planck time $t_{\rm Pl}=\sqrt{\hbar G/c^5}\approx 5.4\times 10^{-44}\,$sec are introduced.
As a benchmark, recent nanoparticle matter-wave interferometry has demonstrated interference of sodium clusters containing more than $7000$ atoms, with masses larger than $170\,{\rm kDa}$~\cite{Pedalino:2025tur}.
This corresponds to
$
    M\simeq 2.8\times10^{-22}\,{\rm kg}
    \simeq 1.3\times10^{-14}m_{\rm Pl}.
$
Taking the logarithmic factor in Eq.~\eqref{eq:Delta_tau_estimate} to be of order unity, one obtains
\begin{equation}
    \Delta\tau
    \sim
    3\times10^{-57}\,{\rm s},
    \qquad
    c\Delta\tau
    \sim
    10^{-48}\,{\rm m}.
\end{equation}
Thus, even for state-of-the-art massive spatial superpositions, the timelike/spacelike-superposed region is extremely small.
Current matter-wave experiments are still many orders of magnitude below the Planck mass.
Progress in mechanical systems is nevertheless encouraging: Schrödinger-cat states of motion have been prepared in a resonator with an effective mass of $16.2\,\mu{\rm g}\approx 0.74m_{\rm Pl}$~\cite{Bild:2022ues}. 
Although this is not the spatial superposition of a localized source assumed here, it suggests that quantum control of near-Planck-mass mechanical systems is becoming realistic. 
Once spatial superpositions with $M\gtrsim m_{\rm Pl}$ become available, Eq.~\eqref{eq:Delta_tau_estimate} shows that $\Delta\tau$ can exceed the Planck time.

%%%%%%%%%%%%%%%%%%%%%%%%%%%%%%%%%%%%%%%%%%%%%%%%%%%
\section{Light-cone Smearing and UV cutoff}
\label{sec:reduced_wightman_smearing}
%%%%%%%%%%%%%%%%%%%%%%%%%%%%%%%%%%%%%%%%%%%%%%%%%%%

In this section we ask what is seen if the source particle is not observed and only the scalar field is probed.
Starting from the corrected Wightman function \eqref{eq:wightman_shifted_sing_sec3}, we take the expectation value over the source state and obtain the reduced Wightman function.
We will see that the quantum Shapiro delay $\hat\tau_{\bm n}$ smears the light-cone singularity in position space, while in Fourier space it acts as an effective UV cutoff in the null direction.
A similar light-cone smearing was discussed for quantized gravitons~
\cite{Ford:1994cr,Ford:1996qc,Matsui:2026ppg}, while here it is induced by the quantum matter source. 

When the source is not observed, the singular part of the reduced Wightman function is obtained by averaging \eqref{eq:wightman_shifted_sing_sec3} over the source state,
\begin{equation}
    \mathcal{G}_{{\rm sing}}^+(u)\equiv 
    \left\langle\hat G_{{\rm sing}}^+(u)\right\rangle_S
    =
    -\frac{1}{8\pi^2 r}
    \left\langle
        \frac{1}{
            u-\hat\tau-i\epsilon
        }
    \right\rangle_S
    =
    -\frac{1}{8\pi^2 r}
    \int \dd\xi\,
    \frac{
        P_{\tau}(\xi)
    }{
        u-\xi-i\epsilon
    } .
    \label{eq:reduced_wightman_convolution}
\end{equation}
Here we have introduced the probability density distribution of the quantum Shapiro delay,
$
    P_{\tau}(\xi)
    =
    \left\langle
        \delta\!\left(
            \xi-\hat\tau
        \right)
    \right\rangle_S
$.
For notational simplicity, we suppress the direction label $\bm n$ on $\hat\tau$ in this section.
If the source state is sharply localized in an eigenstate of $\hat\tau$, then $P_{\tau}(\xi)$ becomes a delta function and the light cone is simply shifted by a $c$-number.
If, on the other hand, the source state has quantum fluctuations, the singularity is no longer localized at a single point; it is smeared over the width of the Shapiro-delay distribution.

To make this explicit, let us define the mean and variance of the quantum Shapiro delay by
\begin{equation}
    \bar\tau 
    \equiv
    \left\langle
        \hat\tau 
    \right\rangle_S,
    \qquad
    \sigma_{\tau}^2
    \equiv
    \left\langle
        \delta\hat \tau^2
    \right\rangle_S\,,
    \qquad \left(\delta\hat\tau\equiv \hat\tau -\bar\tau\right).
    \label{eq:Xi_mean_variance}
\end{equation}
For a Gaussian distribution of $\hat\tau$, the reduced Wightman function can be evaluated analytically:
\begin{align}
    \mathcal{G}_{{\rm sing}}^+(u)
    &=
    -\frac{1}{8\pi^2 r}
    \int
    \frac{\dd\xi}{u-\xi-i\epsilon}
    \frac{
        \exp\!\left[
            -
                \left(
                    \xi-\bar\tau 
                \right)^2
            /(
                2\sigma_{\tau}^2)
        \right]
    }{
    \sqrt{2\pi}\sigma_{\tau}    
    }\,,
    \notag\\
    &=
-\frac{\sqrt{2}}{8\pi^2 r \sigma_{\tau}}
    \left[
        D(z)
        +
        i
        \frac{\sqrt{\pi}}{2}
        e^{-z^2}
    \right],
    \qquad
    \left(z
    \equiv
    \frac{
        u-\bar\tau
    }{
        \sqrt{2} \sigma_{\tau}
    }\right)\,,
    \notag\\
    &\xrightarrow[]{z\ll 1}
    -\frac{\sqrt{2}}{8\pi^2 r \sigma_{\tau}}
    \left[
        i
        \frac{\sqrt{\pi}}{2}
        +z+\mathcal{O}(z^2)
    \right].
    \label{eq:Gaussian_smeared_wightman}
\end{align}
Here
$
    D(x)
    =
    e^{-x^2}
    \int_0^x
    \dd s\,
    e^{s^2}
$
is the Dawson function.
This expression shows explicitly that the singularity that would have been located on the classically shifted light cone $u=\bar\tau$ is rendered finite by the quantum fluctuation $\delta\hat\tau$ of the Shapiro delay.
%On the light cone, the singular and real part of the original Wightman function is replaced by a finite imaginary contribution of order $1/\sigma_\tau$.
This smearing can be understood as a decoherence effect: the scalar wavefront becomes entangled with the source through the quantum Shapiro delay, and tracing out the source causes dephasing among branches with different delays. The sharp light-cone singularity is thereby washed out and replaced by a smooth distribution of width $\sigma_\tau$.

The same effect can be viewed in Fourier space as the emergence of an effective UV cutoff.
Fourier transforming \eqref{eq:reduced_wightman_convolution}, we obtain
\begin{equation}
    \mathcal{G}_{{\rm sing}}^+(u)
    =
    -\frac{i}{8\pi^2 r}
    \int_0^\infty
    \dd k\,
    e^{-ik(u-\bar\tau -i\epsilon)}
    \left\langle
        e^{ik\delta\hat\tau }
    \right\rangle_S .
    \label{eq:reduced_wightman_fourier}
\end{equation}
For a Gaussian distribution,
$
    \left\langle
        e^{ik\delta\hat\tau }
    \right\rangle_S
    =
    \exp\!\left[
        -\frac12
        k^2\sigma_{\tau}^2
    \right]
$.
Therefore the contribution of modes with
\begin{equation}
    k
    \gtrsim
    k_{\rm cut}
    \equiv
     \sigma_{\tau}^{-1}
    \label{eq:effective_null_cutoff}
\end{equation}
is exponentially suppressed.
Thus, in the reduced description in which the source is not observed, the quantum fluctuation of the Shapiro delay produces an effective UV cutoff in the null direction.

A comment on the range of validity is in order.
Although \eqref{eq:reduced_wightman_fourier} is written as the Fourier representation of
$1/(u-\hat\tau-i\epsilon)$, the shifted form itself was inferred from a first-order calculation.
If it were treated strictly as a fixed-order Born expansion, one might require $k\hat\tau\ll1$; indeed, for $k\sigma_\tau\gtrsim1$, the higher powers in
$e^{ik\hat\tau}=1+ik\hat\tau+\cdots$
become important.
Here we instead interpret \eqref{eq:reduced_wightman_fourier} as a resummation of the leading eikonal phase.
A high-frequency null mode propagating through a weak potential acquires the phase shift
$e^{-iku}\to e^{-ik(u-\hat\tau)}$, which is kept to all orders in $k\hat\tau$, while ray bending, amplitude corrections, and multiple scattering remain subleading.
Thus, for $|\Phi_g|\ll1$ and wavelengths shorter than the scale over which the potential varies, the suppression at $k\sigma_\tau\sim1$ should be viewed not as a breakdown of perturbation theory, but as physical dephasing among branches with different Shapiro delays.

This smearing mechanism does not remove the coincidence-limit divergence of the Wightman function~\cite{Ford:1994cr}.
The calculation in this section concerns the singularity near the light cone connecting two points separated by a finite distance $r$, namely $u=t-r\simeq0$ but $r\not\simeq0$.
By contrast, the coincidence limit is obtained by taking $t\to0$ and $r\to0$.
In that limit the length of the ray itself shrinks to zero, and hence the quantum Shapiro delay integrated along the ray also vanishes,
$
    \hat\tau
    =
    -2
    \int_0^r
    \dd\lambda\,
    \Phi_g
    =
    \mathcal{O}(r)
    \to0
$.
Thus the ordinary short-distance divergence of the Wightman function remains in the coincidence limit.
The quantum fluctuation of the Shapiro delay therefore does not replace or remove the Hadamard singularity associated with local composite operators.

Actually, the present effect should be viewed as complementary to Hadamard renormalization widely used in curved-spacetime QFT.
Hadamard renormalization removes the state-independent short-distance singularity in the coincidence limit, but it does not by itself eliminate the sensitivity of light-cone correlations to arbitrarily large null momenta.
That residual UV sensitivity is precisely what is exposed by the light-cone singularity of the Wightman function at finite separation.
Our result shows that, once the quantum source is traced out, this light-cone UV sensitivity is physically suppressed: the quantum fluctuation in the source-induced Shapiro delay provides an effective cutoff at
$k_{\rm cut}\sim\sigma_\tau^{-1}$.
In this sense, the quantum uncertainty of the causal boundary regularizes the finite-distance light-cone singularity without modifying the usual Hadamard structure at coincidence.

%%%%%%%%%%%%%%%%%%%%%%%%%%%%%%%%%%%%%%%%%%%%%%%%%%%
\section{Summary and Discussion}
\label{sec:sum_dic}
%%%%%%%%%%%%%%%%%%%%%%%%%%%%%%%%%%%%%%%%%%%%%%%%%%%

In this paper, we have examined how a quantum matter source affects the causal structure seen by a relativistic quantum field.  We considered a simple weak-field system: a massless scalar field coupled to the Newtonian gravitational potential of a nonrelativistic quantum particle.  In this setting, the Shapiro time delay becomes an operator on the source Hilbert space, and the light-cone singularities of the scalar commutator and Wightman function are displaced by this operator-valued delay.  This has several consequences.  Different causal-boundary shifts need not commute, so the causal structure cannot in general be specified as a single sharp geometry.  A spatial superposition of the source can also make the causal relation between two fixed spacetime points branch dependent.  Finally, when the source is traced out, fluctuations of the delay smear the Wightman light-cone singularity and appear as an effective UV cutoff along the null direction.  These results show that the quantum nature of matter can be inherited by the causal structure itself, even before introducing propagating quantum gravitational degrees of freedom.

A natural next step is to go beyond the singular structure of vacuum correlation functions and follow the quantum state of the scalar field itself.  In the present work, the scalar field was used mainly as a probe of the operator-valued causal boundary.  However, a quantum source can also dress and excite the field in a branch-dependent way.  For a source prepared in a spatial superposition, the resulting field states may get entangled with the source state.  Studying this state-level structure would clarify how different source branches produce distinguishable scalar-field states, and how this entanglement leads to decoherence of the source. It would also help sharpen the relation to indefinite causal order~\cite{Oreshkov:2011er,Zych:2017tau}. The effect found here is not a superposition of the two temporal orders, but rather a branch dependence of whether two events are timelike connected or spacelike separated.  In this sense, it is an indefinite causal relation.  Although experimentally challenging as discussed at the end of Sec.~\ref{sec:causal_relation_superposition}, such a phenomenon could in principle provide a witness of the quantum nature of Newtonian gravity, complementary to the proposed tests of gravity-induced entanglement~\cite{Bose:2017nin,Marletto:2017kzi}.

A more speculative direction is to ask how far this idea can be extended beyond the weak-field regime.  Our calculation is based on a Newtonian potential sourced by a nonrelativistic quantum particle, and therefore cannot be directly applied to strong gravity.  
Nevertheless, if the basic lesson survives more generally, namely that causal boundaries become quantum when the matter sourcing the geometry is quantum, it may have implications for horizon physics. 
In black-hole spacetimes, the horizon is itself a causal boundary.  
One may then ask if a horizon sourced by quantum matter could have quantum fluctuations and be branch-dependent. Such a possibility could blur the separation between the interior and exterior regions of a black hole, affect the mode decomposition of quantum fields near the horizon, and thereby modify the standard QFT description underlying Hawking radiation. A consequence of the ambiguity in the definition of the black hole horizon was discussed in~\cite{Chen:2015gux,Chen:2016nks}.
Similar questions arise for cosmological horizons, especially during inflation, where horizon crossing plays a central role in the generation of quantum fluctuations. These questions lie well outside the validity of the present analysis, but they suggest that quantum causal boundaries could add a new layer to our understanding of horizons in both black-hole physics and cosmology.

\appendix
%%%%%%%%%%%%%%%%%%%%%%%%%%%%%%%%%%%%%%%%%%%%%%%%%%%
\section{Gravitational potential correction to the commutator}
\label{app:commutator_direct_singularity}
%%%%%%%%%%%%%%%%%%%%%%%%%%%%%%%%%%%%%%%%%%%%%%%%%%%

In this appendix we derive the light-cone singularity quoted in
Eq.~\eqref{eq:commutator_shift_expanded}. 
Using \eqref{eq:interaction_picture_Hamiltonian} and
\eqref{eq:micro_3d_phiH_first_order}, the first-order correction to the
commutator \eqref{eq:commutator_def_sec3} is
\begin{align}
    \delta\Delta(t,\bm x;0,\bm 0)
    &=
    i
    \int_0^t d\tau\,
    \left[
        \left[
            \hat H_{{\rm int},I}(\tau),
            \hat\phi_I(t,\bm x)
        \right],
        \hat\phi_I(0,\bm 0)
    \right]
    \notag\\
    &=
    4i
    \int_0^t d\tau
    \int d^3z\,
    \Phi_g\!\left(\bm z-\hat{\bm q}_I(\tau)\right)
    \partial_\tau
    \Delta_0(\tau,\bm z;t,\bm x)
    \partial_\tau
    \Delta_0(\tau,\bm z;0,\bm 0),
    \label{eq:app_deltaDelta_basic}
\end{align}
where
\(\Delta_0(x;y)=[\hat\phi_I(x),\hat\phi_I(y)]\).
For \(0<\tau<t\), we write
\begin{equation}
    \Delta_0(\tau,\bm z;t,\bm x)
    =
    i\mathcal D_1,
    \qquad
    \Delta_0(\tau,\bm z;0,\bm 0)
    =
    -i\mathcal D_0,
    \label{eq:app_Delta_D_def}
\end{equation}
with
\begin{equation}
    \mathcal D_1
    =
    \frac{\delta(t-\tau-R)}{4\pi R},
    \qquad
    \mathcal D_0
    =
    \frac{\delta(\tau-\rho)}{4\pi \rho},
    \qquad
    \rho=|\bm z|,
    \quad
    R=|\bm x-\bm z|.
    \label{eq:app_D1_D0_def}
\end{equation}
Thus the Born correction is supported on broken null paths
\((0,\bm 0)\to(\tau,\bm z)\to(t,\bm x)\).

Since
\(\partial_\tau\mathcal D_1=-\partial_t\mathcal D_1\), and after integrating
by parts in \(\tau\), Eq.~\eqref{eq:app_deltaDelta_basic} becomes
\begin{equation}
    \delta\Delta(t,\bm x;0,\bm 0)
    =
    -4i\partial_t^2 I_\Phi(t,\bm x)
    +
    4i\partial_t J_\Phi(t,\bm x),
    \label{eq:app_deltaDelta_IJ}
\end{equation}
where
\begin{align}
    I_\Phi(t,\bm x)
    &=
    \int d^3z\,
    \frac{
        \Phi_g\!\left(\bm z-\hat{\bm q}_I(\rho)\right)
        \delta(t-S(\bm z))
    }{
        16\pi^2 \rho R
    },
    \label{eq:app_Iphi_def}
    \\
    J_\Phi(t,\bm x)
    &=
    \int d^3z\,
    \frac{
        \dot{\hat\Phi}_I(\rho,\bm z)
        \delta(t-S(\bm z))
    }{
        16\pi^2 \rho R
    },
    \qquad
    S(\bm z)=\rho+R,
    \label{eq:app_Jphi_def}
\end{align}
and
$
    \dot{\hat\Phi}_I(\tau,\bm z)
    \equiv
    \frac{d}{d\tau}
    \Phi_g\!\left(\bm z-\hat{\bm q}_I(\tau)\right).
$
The boundary terms vanish because they contain
\(\mathcal D_1\) or \(\mathcal D_0\) at zero time separation; for example,
\(\int d^3r\,\delta(|\bm r|)/(4\pi|\bm r|)\,f(\bm r)
\propto \int_0^\infty d|\bm r|\,|\bm r|\,\delta(|\bm r|)=0\).

We now extract the singular behavior near the flat-space light cone using
the notation \eqref{eq:r_n_u_def}. 
Near the unperturbed ray, set
\begin{equation}
    \bm z
    =
    \lambda\bm n+\bm y,
    \qquad
    \bm n\cdot\bm y=0,
    \qquad
    0<\lambda<r .
    \label{eq:app_near_ray_coordinates}
\end{equation}
Here \(\bm y\) is the two-dimensional transverse displacement from the
straight null ray connecting \(\bm 0\) and \(r\bm n\).
We expand the integrands of $I_\Phi$ and $J_\Phi$ with respect to $y$ using
\begin{equation}
    S(\bm z)
    =
    r
    +
    \frac{r}{2\lambda(r-\lambda)}
    \bm y^2
    +\cdots,
    \qquad
    \rho R
    =
    \lambda(r-\lambda)+\cdots .
    \label{eq:app_S_near_ray}
\end{equation}
Substituting \eqref{eq:app_S_near_ray} into \eqref{eq:app_Iphi_def} and performing the $y$ integrals, we find
\begin{align}
    I_\Phi(t,r\bm n)
    &=
    \frac{\Theta(u)}{8\pi r}
    \int_0^r d\lambda\,
    \Phi_g\!\left(
        \lambda\bm n-\hat{\bm q}_I(\lambda)
    \right)
    +
    \mathcal O\!\left(u\Theta(u)\right).
    \label{eq:app_Iphi_lightcone}
    \\
    J_\Phi(t,r\bm n)
    &=
    \frac{\Theta(u)}{8\pi r}
    \int_0^r d\lambda\,
    \dot{\hat\Phi}_I(\lambda,\lambda\bm n)
    +
    \mathcal O\!\left(u\Theta(u)\right).
    \label{eq:app_Jphi_lightcone}
\end{align}
The term \(J_\Phi\), as well as the subleading
\(\mathcal O(u\Theta(u))\) part of \(I_\Phi\), generates only terms
proportional to \(\delta(u)\) in \(\delta\Delta\). These terms modify the
amplitude of the direct wavefront but not its position and hence we ignore them in this calculation.

The position of the light-cone singularity is therefore determined by
the leading term in \(I_\Phi\). From \eqref{eq:app_deltaDelta_IJ} and
\eqref{eq:app_Iphi_lightcone}, we obtain
\begin{equation}
    \delta\Delta_{\rm shift}(t,r\bm n;0,\bm 0)
    =
    -\frac{i}{2\pi r}
    \left[
        \int_0^r d\lambda\,
        \Phi_g\!\left(
            \lambda\bm n-\hat{\bm q}_I(\lambda)
        \right)
    \right]
    \delta'(u)
    =
    \frac{i}{4\pi r}
    \hat\tau_{\bm n}\,
    \delta'(u),
    \label{eq:app_deltaDelta_shift_tau}
\end{equation}
where $\hat\tau_{\bm n}$ is defined in \eqref{eq:quantum_Xi_def}.
Together with the free commutator term, this gives Eq.~\eqref{eq:commutator_shift_expanded}.

%%%%%%%%%%%%%%%%%%%%%%%%%%%%%%%%%%%%%%%%%%%%%%%%%%%
\section{Gravitational potential correction to the Wightman function}
\label{app:wightman_direct_singularity}
%%%%%%%%%%%%%%%%%%%%%%%%%%%%%%%%%%%%%%%%%%%%%%%%%%%

We next derive the light-cone singularity quoted in
Eq.~\eqref{eq:wightman_shifted_sing_sec3}. 
The calculation is parallel to Appendix~\ref{app:commutator_direct_singularity}.
The relevant contribution comes from the insertion point lying between the
two external points, \(0<\tau<t\):
\begin{align}
    \delta\hat G^+(t,\bm x;0,\bm 0)
    &=
    i
    \int_0^t d\tau\,
    \bra{0_\phi}
    \left[
        \hat H_{{\rm int},I}(\tau),
        \hat\phi_I(t,\bm x)
    \right]
    \hat\phi_I(0,\bm 0)
    \ket{0_\phi}
    \notag\\
    &=
    4i
    \int_0^t d\tau
    \int d^3z\,
    \Phi_g\!\left(\bm z-\hat{\bm q}_I(\tau)\right)
    \partial_\tau\Delta_0(\tau,\bm z;t,\bm x)
    \partial_\tau G_0^+(\tau,\bm z;0,\bm 0).
    \label{eq:app_wightman_basic}
\end{align}
The remaining first-order terms from the integral over $\tau<0$ do not affect the light-cone shift. Instead, they describe the dressing of the vacuum correlation by the gravitational potential in the common past of
the two external points.

For \(0<\tau<t\), the singular factors near the broken null path are
\begin{equation}
    \Delta_0(\tau,\bm z;t,\bm x)
    =
    i\frac{\delta(t-\tau-R)}{4\pi R},
    \qquad
    G_{0,{\rm sing}}^+(\tau,\bm z;0,\bm 0)
    =
    -\frac{1}{8\pi^2\rho}
    \frac{1}{\tau-\rho-i\epsilon},
    \label{eq:app_wightman_free_sing}
\end{equation}
where $\rho$ and $R$ are defined in \eqref{eq:app_D1_D0_def}.
Substituting these expressions into \eqref{eq:app_wightman_basic}
and using integral by part for $\partial_\tau \Delta_0$,
one obtains
\begin{equation}
    \delta\hat G^+_{\rm sing}
    =
    -\frac{1}{4\pi^3}
    \int d^3z\,
    \frac{
        \Phi_g\!\left(\bm z-\hat{\bm q}_I(t-R)\right)
    }{
        \rho R
    }
    \frac{1}{
        \left(t-\rho-R-i\epsilon\right)^3
    }
    +
    \cdots ,
    \label{eq:app_wightman_after_tau}
\end{equation}
where the omitted terms contain \(\dot{\hat\Phi}_I\) and are less singular
near the light cone.
We now expand the integral near the unperturbed ray as in
Appendix~\ref{app:commutator_direct_singularity}. 
Using \eqref{eq:app_S_near_ray} we find
\begin{align}
    \delta\hat G^+_{\rm sing}
    &=
    -\frac{1}{4\pi^3}
    \int_0^r d\lambda\,
    \frac{
        \Phi_g\!\left(\lambda\bm n-\hat{\bm q}_I(\lambda)\right)
    }{
        \lambda(r-\lambda)
    }
    \int d^2y\,
    \frac{1}{
        \left[
            u-i\epsilon
            -
            \dfrac{r}{2\lambda(r-\lambda)}
            \bm y^2
        \right]^3
    }
    +
    \cdots ,
    \notag\\
    &=
    \frac{1}{4\pi^2 r}
    \left[
        \int_0^r d\lambda\,
        \Phi_g\!\left(
            \lambda\bm n-\hat{\bm q}_I(\lambda)
        \right)
    \right]
    \frac{1}{
        \left(u-i\epsilon\right)^2
    }
    +
    \cdots ,
    \notag\\
    &=
    -\frac{1}{8\pi^2 r}
    \frac{
        \hat\tau_{\bm n}
    }{
        \left(u-i\epsilon\right)^2
    }
    +
    \cdots .
    \label{eq:app_wightman_shift_tau}
\end{align}
Adding the free singular part of the Wightman function then gives
Eq.~\eqref{eq:wightman_shifted_sing_sec3}. 
Thus the Wightman light-cone singularity is shifted by the same
operator-valued Shapiro delay as the commutator.

%%%%%%%%%%%%%%%%%%%%%%%%%%%%%%%%%%%%%%%
\acknowledgments
%%%%%%%%%%%%%%%%%%%%%%%%%%%%%%%%%%%%%%%

We thank 
Anamaria Hell, Youka Kaku, Hiroki Matsui, Akira Matsumura, Shinji Mukohyama, Takahiro Tanaka
for useful discussions.
This work was supported in part by JSPS KAKENHI Grant Nos. JP26H00407 and JP23K03424 (T.F.), and JP24K00624 (M.S.).

% Bibliography

%% [A] Recommended: using JHEP.bst file
%% \bibliographystyle{JHEP}
%% \bibliography{biblio.bib}

%% or
%% [B] Manual formatting (see below)
%% (i) We suggest to always provide author, title and journal data or doi:
%% in short all the informations that clearly identify a document.
%% (ii) please avoid comments such as "For a review'', "For some examples",
%% "and references therein" or move them in the text. In general, please leave only references in the bibliography and move all
%% accessory text in footnotes.
%% (iii) Also, please have only one work for each \bibitem.

\end{document}